\documentclass[%
 aip,jcp,       
 amsmath,amssymb,
 reprint,       
]{revtex4-1}

\usepackage[pdftex]{graphicx}
\usepackage{dcolumn}
\usepackage{bm}

\usepackage[utf8]{inputenc}
\usepackage[T1]{fontenc}
\usepackage{mathptmx}
\usepackage{etoolbox}

\usepackage{xcolor}
\usepackage{hyperref}
\usepackage{xspace}

\usepackage{float}

\makeatletter
\def\@email#1#2{%
 \endgroup
 \patchcmd{\titleblock@produce}
  {\frontmatter@RRAPformat}
  {\frontmatter@RRAPformat{\produce@RRAP{*#1\href{mailto:#2}{#2}}}\frontmatter@RRAPformat}
  {}{}
}%
\makeatother

\begin{document}

\title{An expandable kinetic Monte Carlo platform for modelling electron transport through chiral molecules}

\author{Silvia Gim\'enez-Santamarina}
\email{silvia.m.gimenez@uv.es, gimenezs@tcd.ie}
\affiliation{Instituto de Ciencia Molecular (ICMol), Universitat de Val\`encia, Paterna, Spain}
\affiliation{School of Physics, Trinity College Dublin, Dublin, Ireland}

\author{Gerliz M. Guti\'errez-Finol}
\affiliation{Instituto de Ciencia Molecular (ICMol), Universitat de Val\`encia, Paterna, Spain}

\author{Andr\'es Mora-Mart\'inez}
\affiliation{Instituto de Ciencia Molecular (ICMol), Universitat de Val\`encia, Paterna, Spain}

\author{Alejandro Gaita-Ari\~no}
\email{gaita@uv.es}
\affiliation{Instituto de Ciencia Molecular (ICMol), Universitat de Val\`encia, Paterna, Spain}

\date{\today}

\begin{abstract}

Chirality from molecular structures interacts with the spin angular momentum of electrons and photons giving rise to a variety of interesting phenomena. Among these, the observation of spin selective transport at room temperature is particularly attractive for the development of functional spintronic devices. During the past twenty five years, two effects have attracted considerable experimental and theoretical attention: electric Magnetochiral Anisotropy (eMChA) and Chirality Induced Spin Selectivity (CISS). In spite of the large body of work devoted to these phenomena, there is still no clear agreement on their microscopic origin(s). It even remains an open question whether eMChA and CISS arise from fundamentally different mechanisms or whether they are different experimental manifestations of the same underlying microscopic effects. In this work, we have developed the core of an efficient kinetic Monte Carlo code for the modeling of electron transport under an applied voltage, where the $\alpha$ and $\beta$ spin channels are treated independently. Each transport channel is characterized through its intrinsic electron mobility and an effective coupling between charge motion, spin, and chirality. This framework makes it possible to quantify the spin filtering that emerges from the interplay of these properties. From a set of simple rules emerges a voltage- and field-dependent effect that vanishes at low bias and shows the asymmetry between positive and negative voltages that is typically reported in electrical magnetochiral anisotropy experiments. We then move on to relate the internal parameters of our code with parameterization that has been used to describe the eMChA effect. 
\end{abstract}


\maketitle

\section{Introduction}

\subsection{Spintronics Without Magnets}

Spintronics is based on the idea that the spin of the electron can be used together with its charge to transport, process and store information. Treating spin as an additional degree of freedom, opens the door to new devices that go beyond conventional electronics, such as non-volatile memories and low power logic elements. The discovery of spin-dependent transport in magnetic multilayers established the foundations of modern spintronics by demonstrating that electron transport can be manipulated through spin-dependent scattering effects, thereby revealing the strong potential of spin-selective transport for information technologies. \cite{Wolf2001, Zutic2004}

Most spintronic devices, however, still rely on ferromagnetic materials, and often on external magnetic fields, to generate, manipulate or detect spin currents. These requirements introduce several practical limitations. The presence of ferromagnets complicate miniaturization as they can produce stray fields that interfere with neighboring components complicating the integration of the device. \cite{Hirohata2020} The difficulties become even more severe when the objective is integration with molecular, organic or biological systems. These limitations have motivated the search for alternative mechanisms capable of controlling spin without the need of magnetic order, and preferably, that work at room, or nearly-room temperature. \cite{Gupta2024, Bloom2024, FormentAliaga2022} 

Several non magnetic mechanisms have been explored in recent years. For instance, materials with strong spin-orbit coupling and structural asymmetry can generate spin polarization without the use of ferromagnets. \cite{Manchon2019} In this scenario, chirality plays an intriguing role.

Chiral molecules lack mirror symmetry, and this structural property has been shown to influence electron transport in a way that depends on spin. Experiments demonstrate that electrons transmitted through chiral systems can become spin polarized even when the material itself contains no magnetic elements.\cite{Naaman2015} This observation suggests that spin control may be achieved through molecular structure alone,  opening a path toward chiral systems as potential platforms for developing lighter and more versatile spintronic devices, with significant potential application in quantum technologies. \cite{Shang2022, Chiesa2025}


\subsection{Chirality and the Origin of Spin Selectivity}

As mentioned above, molecular chirality refers to the absence of mirror and inversion symmetry in a structure, meaning that it cannot be superimposed on its mirror image. This structural property is not only geometric, but it has also been shown to have direct consequences on the optical and electrical properties of molecular systems. 


When spatial inversion symmetry is broken, couplings can arise between quantities that would otherwise remain independent. In chiral systems, this allows a connection between the linear momentum of an electron and its spin angular momentum. As a result, the spin orientation of an electron moving through a chiral potential becomes correlated with its direction of motion. \cite{Naaman2012, Naaman2016}


One of the best known consequences that result from the interaction of chirality with fundamental quantum degrees of freedom, including charge and spin, is the Chirality-Induced Spin Selectivity (CISS) effect. Here, an initially unpolarized current becomes spin polarized after being transmitted through a chiral material, with a polarization whose sign depends on the handedness of the molecule and on the direction of transport. The effect was first observed in the transmission of photoelectrons through organized films of chiral molecules. \cite{Naaman1999} It has since been reported in self assembled monolayers of double stranded DNA, \cite{Gohler2011} in helicenes, \cite{Kettner2018} in biological membranes, \cite{Mishra2013} in chiral peptides and metallopeptides, \cite{TorresCavanillas2020} in helicoidal metal organic frameworks \cite{HuiziRayo2020} and in organic radical monolayers, \cite{Giaconi2024} among many other systems. \cite{Naaman2019, Gupta2024, Bloom2024} The measurements are equally varied, and include magnetic conductive atomic force microscopy, photoemission spectroscopy, magnetoresistance and electrochemical techniques.

Two aspects of these studies are particularly relevant. First, the reported spin selectivities are large, with values of 50\%, and even 70\% or 100\% in the presence of a paramagnetic ion. \cite{TorresCavanillas2020, HuiziRayo2020} Second, they are obtained at room temperature in layers that carry no magnetic order of their own, so it is the symmetry of the molecule, and not a magnetic material, which determines the spin selectivity. \cite{Naaman2012, Naaman2019, Bloom2024} The robustness of this effect is one of its most remarkable aspects. Spin polarization appears under ambient conditions in materials that are not magnetic, in contrast with conventional spin dependent phenomena that typically weaken at high temperature. \cite{Naaman2012, Naaman2016} From a physical perspective, this behavior shows that spin can be controlled through the symmetry properties of matter. The handedness of the molecular structure determines the preferred spin orientation, establishing a direct link between molecular structure and spin transport. 

\subsection{Experimental Landscape: CISS and eMChA}

\begin{figure*}
    \centering
    \includegraphics[width=0.7\textwidth]{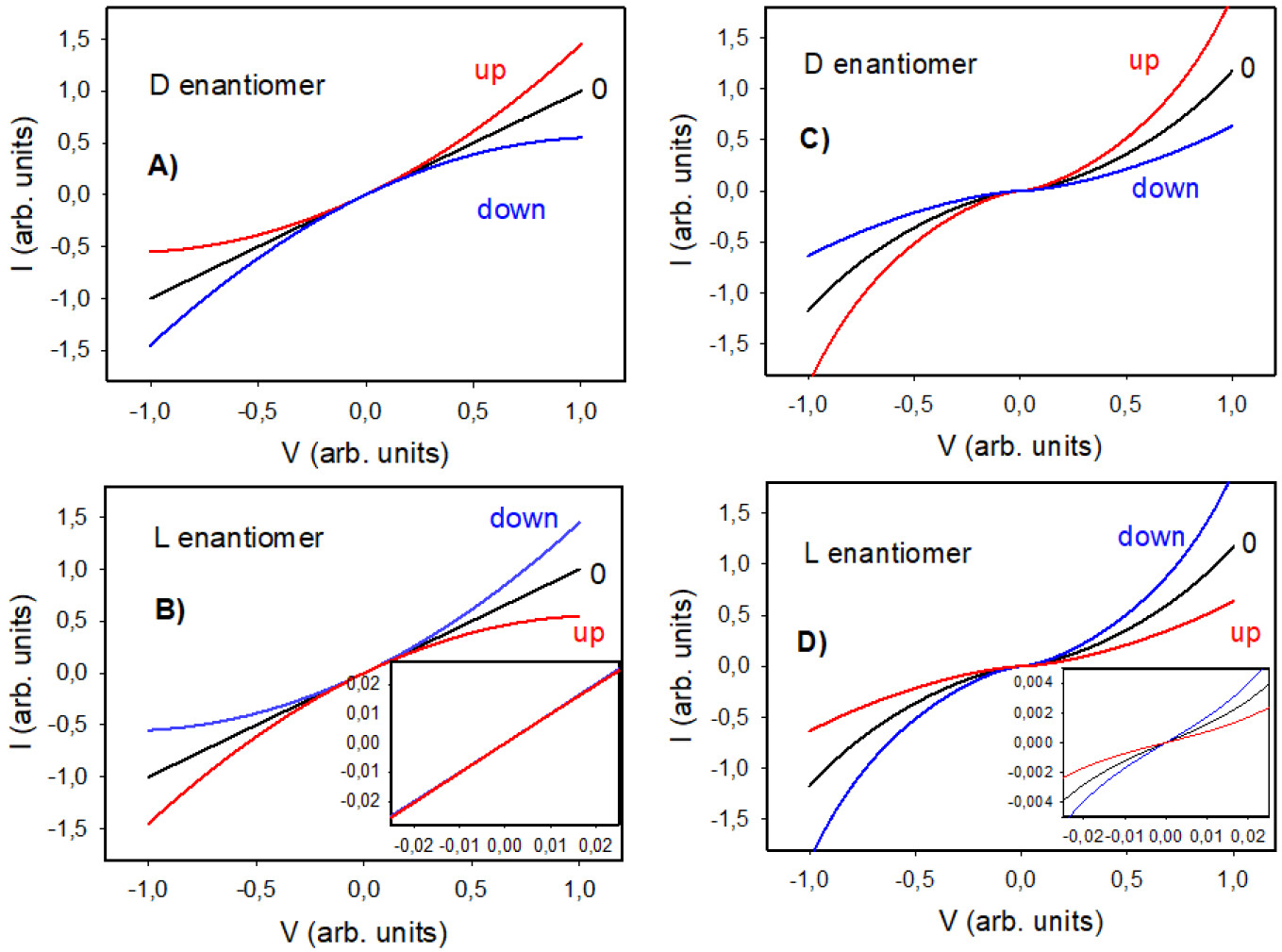}
    \caption{Qualitative scheme comparing the current/voltage response of different spins (up in red, down in blue) passing through different enantiomers (D in upper panels, L in lower panels), in different directions (positive or negative voltages), in different experiments (eMChA in left panels, CISS in right panels). Inset zooms highlight the different behaviors at low voltage. Reproduced with permission from \cite{Rikken2023}}
    \label{fig:Rikken}
\end{figure*}

The experimental discussion of spin dependent transport in chiral systems is usually framed in terms of two closely related effects: CISS, introduced above, and electrical magnetochiral anisotropy (eMChA). Both are understood as consequences of the coupling between electron motion and a chiral electrostatic potential, although  they are identified through different experimental protocols and display distinct transport signatures, as shown in Figure~\ref{fig:Rikken}. \cite{Naaman2019, Rikken2023}

CISS refers to the observation of an initially unpolarized electron current that can become spin polarized after passing through a chiral conductor. The effect depends on the handedness of the molecular structure and on the direction of charge transport, but it does not require an external magnetic field. \cite{Naaman1999, Naaman2012, Naaman2016, Naaman2019, Giaconi2024} In many experiments, the measured spin polarization shows only a weak dependence on the applied voltage and remains robust at room temperature. \cite{Naaman2012, Naaman2019} In a CISS experiment, the quantity of interest is therefore the spin polarization of the transmitted current. Since direct spin-resolved detection is technically challenging, the effect is most often probed indirectly by placing a ferromagnetic contact in the circuit and comparing the current for opposite orientations of its magnetization. Magnetic elements may therefore appear in the experimental setup, but not in the chiral material itself. \cite{Naaman2012}

eMChA, in contrast, is a different measurement and belongs to a wider family of magnetochiral effects. Magnetochiral anisotropy was first observed optically, as a difference in the absorption of light by a chiral medium placed in a magnetic field, \cite{Rikken1997}, it has also been found in electrical transport \cite{Rikken2001, Atzori2021} and even in the propagation of phonons. \cite{Nomura2019} In its electrical version, the resistance of a chiral conductor is measured in the presence of (i) a current and (ii) a magnetic field, and it depends on their relative orientation: reversing either the current or the field changes the resistance, while reversing both leaves it unchanged. Its sign is set by the handedness of the conductor, which is what makes it a chiral effect rather than an ordinary magnetoresistance. \cite{Pop2014} Because the signal is odd in the current, it belongs to the nonlinear part of the transport response and disappears in the linear regime, at low bias. \cite{Rikken2001, Pop2014}

Although CISS and eMChA are usually presented as distinct phenomena, they share a common symmetry origin. Both arise in systems lacking inversion symmetry and are associated with the coupling between electron motion and the chiral potential. Their main differences lie in how the effects are detected and in their dependence on external magnetic fields and applied voltage. In CISS, the spin selectivity is associated with the handedness of the molecular structure and the direction of transport, whereas eMChA requires the simultaneous presence of a current and an external magnetic field and appears as a nonlinear magnetotransport response. Experimentally, both effects have been reported in a wide range of systems, including molecular monolayers, chiral crystals, perovskites, and hybrid interfaces, suggesting that the underlying physics is not restricted to a specific class of materials. \cite{Xu2023, Long2020, Yan2024, Atzori2021, Meng2021}

The use of ferromagnetic contacts in CISS measurements, however, introduces an important experimental consideration. In several commonly used analyses, the extracted spin polarization depends on assumptions about the spin polarization of the injected or detected current. Since real ferromagnetic contacts generally exhibit only partial spin polarization, assuming an ideal spin injector or analyzer can lead to an overestimation of the inferred spin selectivity. Moreover, the current is measured under the same external magnetic field used to polarize the ferromagnetic contact. If CISS and eMChA are indeed distinct phenomena, the measured signal may therefore contain contributions from both effects. This has motivated caution in interpreting the reported values of spin selectivity.

Despite these considerations, some features of the experimental observations remain robust. The observed spin selectivity often depends only weakly on the applied voltage and persists up to room temperature, while the chiral material itself carries no magnetic order. \cite{Naaman2012, Naaman2019, Rikken2023} At the same time, despite the large amount of experimental and theoretical work devoted to these effects, there is still no consensus on a widely accepted microscopic explanation. Proposed models involve spin-orbit coupling, quantum coherence, many-body interactions, decoherence processes, and vibronic effects, among others. Because geometry, symmetry breaking, spin dynamics, and non-equilibrium transport are closely intertwined, it remains unclear whether CISS and eMChA should be regarded as fundamentally different phenomena or as different manifestations of a common chiral spin-transport mechanism operating under different conditions. \cite{Bloom2024, Rikken2023}

\subsection{Overview of Theoretical Models for CISS or eMChA}

Theoretical descriptions of the CISS or eMChA effects generally rely on three main ingredients: spin-orbit coupling, the absence of inversion symmetry associated with chirality, and electron transport under non-equilibrium conditions. What distinguishes the various families of models is how these physical quantities relate, and which of their contributions dominates on each of the effects.

\subsubsection{Single particle and tight binding models}

Early and minimal descriptions represent a chiral molecule as a one dimensional or quasi one dimensional tight binding chain with a helical geometry or twisted orbitals. In these models an effective spin-orbit interaction acts on electrons moving through a chiral potential. Transport is typically analyzed using Landauer theory or non-equilibrium Green function formalisms (details of this method can be found in \cite{Camsari2022}). These models are useful to clarify the symmetry requirements needed to produce spin dependent transmission, and they can generate finite spin polarization. However, they often require carefully tuned parameters and usually predict smaller spin polarization than the values reported in experiments. \cite{Bloom2024, Ghazaryan2020}

\subsubsection{Non equilibrium transport approaches}

More elaborate treatments describe the chiral region as a conductor placed between metallic leads and compute the current and spin polarization using non equilibrium Green function techniques or related transport methods. This framework allows explicit control of parameters such as applied voltage, temperature, and dephasing processes. It also makes it possible to investigate the influence of disorder and molecule contact coupling on the observed signal. Although these approaches provide a detailed description of transport processes, they are computationally demanding and still have difficulty reproducing the large spin polarizations observed experimentally over broad parameter ranges. \cite{Bloom2024}

\subsubsection{Correlated electron and current constrained models}

Another line of research focuses on correlated electron Hamiltonians in which a current is imposed directly through the system. For example, Hubbard chains with twisted orbitals have been studied under a fixed current constraint rather than through explicit leads. These studies show that electron-electron interactions and nonadiabatic vibrational effects can enhance spin polarization, especially in systems that are not half filled or that contain weak bonds. At the same time, they indicate that very large CISS signals are difficult to obtain in purely electronic models without introducing additional physical mechanisms. \cite{Savi2025}

\subsubsection{Vibronic and open quantum system approaches}

A growing number of studies emphasize the role of coupling between electronic motion and molecular vibrations, as well as interactions with the surrounding environment. In these approaches the electron vibration interaction is treated explicitly, often within mixed quantum classical descriptions or open quantum system frameworks. These models suggest that non-equilibrium vibrational dynamics can generate or amplify spin selectivity in chiral structures. Such mechanisms may help explain the robustness of the CISS effect at room temperature and indicate that environmental interactions may play an active role in the phenomenon rather than simply degrading the signal. \cite{Savi2025, Chiesa2025}

\subsubsection{Current status and open questions}

Recent reviews and critical analyses point out that no single theoretical framework is currently able to reproduce simultaneously the magnitude, voltage dependence, and material generality observed in CISS experiments. These studies highlight the possibility that additional mechanisms may need to be considered, including exchange interactions, displacement currents, or more complex couplings between spin and molecular vibrations. \cite{Fransson2025, Bloom2024, Eckvahl2023, Xu2023}

\subsection{Microscopic, frugal, expandable modelling}

Addressing these open questions requires theoretical frameworks that allow microscopic assumptions to be tested efficiently, in a controlled way, while connecting the results directly with experimentally accessible observables. In this work we develop a microscopic phenomenological model based on kinetic Monte Carlo (kMC) methods to describe charge and spin transport under an applied electric potential in chiral molecular structures, with the long term goal of gaining insight into the mechanisms that may govern both CISS and eMChA.

Kinetic Monte Carlo approaches naturally capture non-equilibrium stochastic transport processes on discrete networks while remaining computationally efficient over experimentally relevant time and length scales. Similar methodologies have successfully reproduced key experimental features in other contexts, including magnetic memory effects in nanomagnets \cite{GutierrezFinol2025DAISY} and memristive behavior in solid state ion conduction. \cite{GutierrezFinol2026CupFlow}

In this work we present the core of an efficient kinetic Monte Carlo model that explicitly incorporates spin degrees of freedom, molecular chirality, and a voltage bias. In this initial implementation we demonstrate the minimal functionality of the model and illustrate how effective couplings between charge motion, spin, and helicity give rise to spin dependent transport signatures. The modular structure of the code is designed to facilitate the systematic incorporation of existing and emerging theoretical mechanisms and to enable the reuse of experimental data to benchmark different models against the growing body of experimental evidence.

This paper is organised as follows. First, we will introduce the basics of our modelling of electron transport, and how the model implemented in the present work relates with the CupFlow kinetic Monte Carlo code that was presented recently for ion transport in systems with memristive behavior.\cite{GutierrezFinol2026-CupFlow} Then we will validate this part of the model, by demonstrating that our software recovers the expected behavior according to Ohm's law and the known resistivity equations, which are not implicitly included in the model. Then we will exemplify the simulation of current through a DNA nanowire, showing which parameters would be needed to recover a classic experiment in nanoelectronics.\cite{Fink1999} In the next stage, we will show how we introduce spin in the modelling. Subsequently we will analyze the qualitative behavior of our implementation, in particular the symmetry of the spin-dependent $I$--$V$ response and critically compare it with the shape of the curves commonly obtained for either CISS or eMChA experiments. Finally, we will analyze also the quantitative behavior of our implementation, and extract the spin-chirality interaction parameters $\gamma$ and $\Delta R$ as defined by Pop et al.\cite{Pop2014} We will offer some final remarks and perspectives for further development of the tool we present here.

\section{Results and Discussion}

\subsection{Basic electron transport modelling and relation with the CupFlow code}

\begin{figure*}
    \centering
    \includegraphics[width=.7\linewidth]{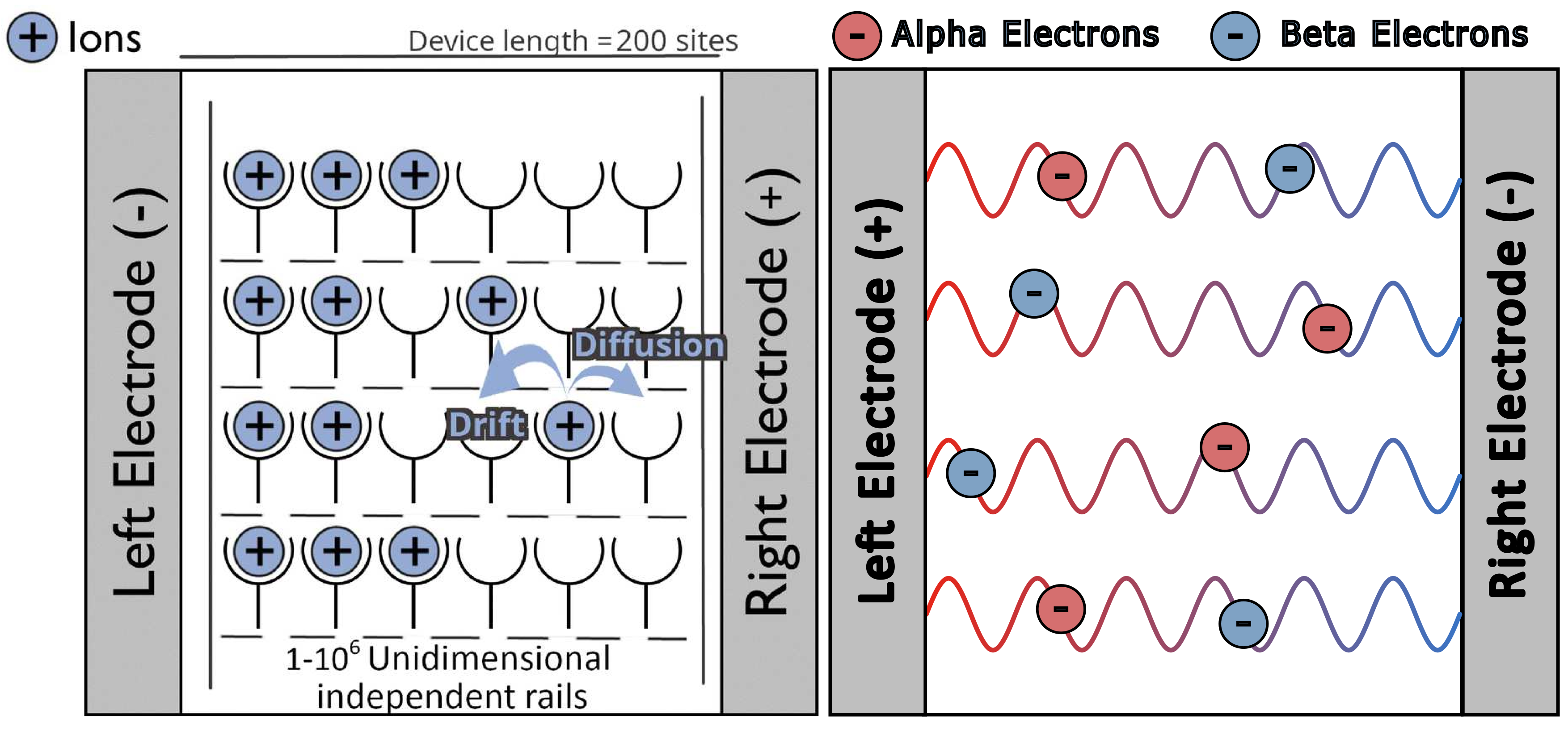}
    \caption{Left: schematic illustration of the kinetic Monte Carlo model used to describe ion transport within the CupFlow code. Ions move between different positions along independent pathways. The movement is stochastic, allowing for a random diffusion, but in presence of an electric voltage the probabilities are biased, resulting in a systematic drift and charge accumulation near one of the electrodes. Reproduced with permission from \cite{GutierrezFinol2026-CupFlow}. Right: illustration showing some of the differences with the current code. First, particles have now negative charge, but more importantly they now have spin (alpha in blue, beta in red). And crucially, the conduction channels favor a spin-dependent direction of movement, which we depict with a color gradient. We only show one case, but the color gradient would be flipped by changing either the chirality of the molecules.}
    \label{fig:CupFlow}
\end{figure*}

The present implementation models electron transport in an ensemble of $N$ identical molecular strands, each hosting a single conducting electron, under the influence of an external voltage applied by electrodes at the two ends of the molecules. The initial position of the electron is exactly at the middle of our partition, therefore demanding an odd number of regions. The applied voltage $V$ and the molecular length $l$ are treated as external parameters. Together with the electron charge $e$, they define a total electrostatic energy drop across each molecule, $E = V \cdot e$.
A linear potential profile is assumed along the molecular backbone, such that an electron displacement over a distance $\Delta_l$ corresponds to an energy change:
\begin{equation}
\Delta E = \frac{\Delta_l}{l} V e
\label{eq:energy_step}
\end{equation}

The intrinsic electron mobility of the molecules at room temperature is parametrized by a characteristic hopping time $\tau$, with the diffusion coefficient $D$ being proportional to the inverse of the time $1/\tau$:
\begin{equation}
   D \propto 1/\tau
\end{equation}

A straightforward extension of the model consists in expressing $\tau$ in terms of an effective energy barrier $U_{\mathrm{eff}}$, an attempt time $\tau_0$, and the temperature $T$, for instance through an Orbach-type activated process:
\begin{equation}
\tau = \tau_0 \exp\left(-\frac{U_{\mathrm{eff}}}{T}\right) .
\label{eq:orbach_tau}
\end{equation}

Within these assumptions, the kinetic Monte Carlo algorithm interrogates, at each time step, whether each electron undergoes a hopping event, following the same strategy previously validated for magnetic relaxation processes in the STOSS and DAISY codes. \cite{gutierrez2023lanthanide,GutierrezFinol2025DAISY} When a hopping event occurs, the direction of electron motion is determined probabilistically, with transition probabilities proportional to the ratio of Boltzmann populations evaluated at the working temperature $T$ and the corresponding energy difference $\Delta E$, in direct analogy with the procedure employed in the CupFlow code for ion transport, see Figure \ref{fig:CupFlow}.\cite{GutierrezFinol2026-CupFlow} This approach represents a natural extension of the methodology originally developed for spin relaxation in STOSS to charge transport under an applied electric field.

In Figure \ref{fig:CupFlow} we represent one of key differences between the two models, concerning the interaction of spin and chirality and its effect on the direction of the movement of the charges. Another key difference between our present model and CupFlow is that the latter was designed to model ion migration within solid-state devices, where continuous electrical current cannot be directly defined and only charge accumulation near the electrodes can be monitored. In the present implementation, we extend the framework to explicitly allow for steady-state current flow. Specifically, when an electron exits the system through the right end of the molecule, it is considered drained, reintroduced at the left end to maintain the total number of charge carriers, and a corresponding event is recorded. Conversely, when an electron exits through the left end, it is reintroduced at the right end, then, a sourced event is registered. The net electron current is defined as the difference between sourced and drained events, normalized by the total simulation time.

\subsection{Validation: recovering Ohm's law and resistivity equations}

The stochastic transport model introduced above does not explicitly incorporate dissipative mechanisms, nor does it impose macroscopic circuit relations by construction. In particular, neither Ohm’s law, $I=V/R$, nor the standard resistivity relationship, $R=\rho l/A$, is imposed at any stage of the simulation. These relations are therefore not built into the model but must instead emerge from the underlying microscopic dynamics. Establishing that the model naturally reproduces the expected linear current--voltage response and the dependence of resistance on molecular length and cross-sectional area constitutes an essential validation step before addressing more complex spin- and chirality-dependent transport effects.

Within the model, charge transport proceeds through stochastic forward and backward electron hops along the molecular backbone, with a slight bias introduced by the applied electric field. This bias favors motion along the field direction and depends on the total voltage drop across the molecule as well as on its length. Despite the simplicity of this mechanism, the resulting steady-state current displays a linear dependence on the applied bias and decreases with increasing molecular length, reproducing the expected macroscopic behavior associated with Ohmic transport.

The magnitude of the current also scales with $N$, the number of parallel molecular strands contributing to transport, which plays the role of an effective cross-sectional area. For each individual molecule, the material resistivity is controlled by the characteristic time associated with individual hopping events, or equivalently by the effective charge diffusivity. For example, let us introduce a diffusion parameter $D=0.5$, corresponding to a 50\% probability of an electron moving from one region to a neighbouring one after a time step of $10^{-17}$~s. At zero voltage, an electron that at $t_0$ is in the $i$-th region will have, at $t=t_0+10^{-17}$~s, a 50\% probability of remaining in the $i$-th region, a 25\% probability of shifting to the $(i+1)$-th region, and another 25\% probability of shifting to the $(i-1)$-th region. Application of a voltage changes the ratio between the two latter probabilities, while leaving the probability of "staying put" constant. The characteristic hopping time therefore determines the effective diffusivity and, consequently, the resistivity of the system.

When the hopping time is modeled as a thermally activated process, the resulting conductivity exhibits an exponential dependence on temperature, as commonly observed in semiconducting systems. In this formulation, the attempt time reflects the choice of the elementary spatial discretization used to represent the molecular structure.

The model can also introduce saturation effects, depending on the combination of temperature, voltage, and spatial partition. Of these parameters, temperature and voltage have a clear physical meaning, whereas the dependence on the spatial partition is a consequence of the numerical discretization. At low temperatures, high voltages, and/or coarse spatial partitions, the simulated current--voltage response deviates from linearity and begins to saturate. In the present work, we choose spatial partitions that keep the system within the linear-response regime. Although this saturation is therefore not central to the present discussion, its effect is monitored throughout the simulations.

Taken together, these observations demonstrate that standard resistive transport behavior emerges naturally from the stochastic electron dynamics implemented in the model, even in the absence of explicitly imposed circuit-level relations.

In contrast, what the current version of the model does not reproduce is the buildup voltage often displayed by molecules with discrete energy levels between two metallic electrodes, where one typically finds zero current until a threshold voltage is reached and the linear response begins. In the present model, however, the electrodes do not exist explicitly. When an electron moves to the right from the rightmost position, it simply exits the molecule, while simultaneously another equivalent electron enters at the leftmost position. In other words, the model assumes that whatever the energy is of the orbitals that describe the delocalized electrons, the electron will always find both a conduction band able to accept electrons and a valence band able to provide electrons, as needed.



\subsection{Simulating current through a DNA nanowire}

\begin{figure*}
    \centering
    \includegraphics[width=.7\linewidth]{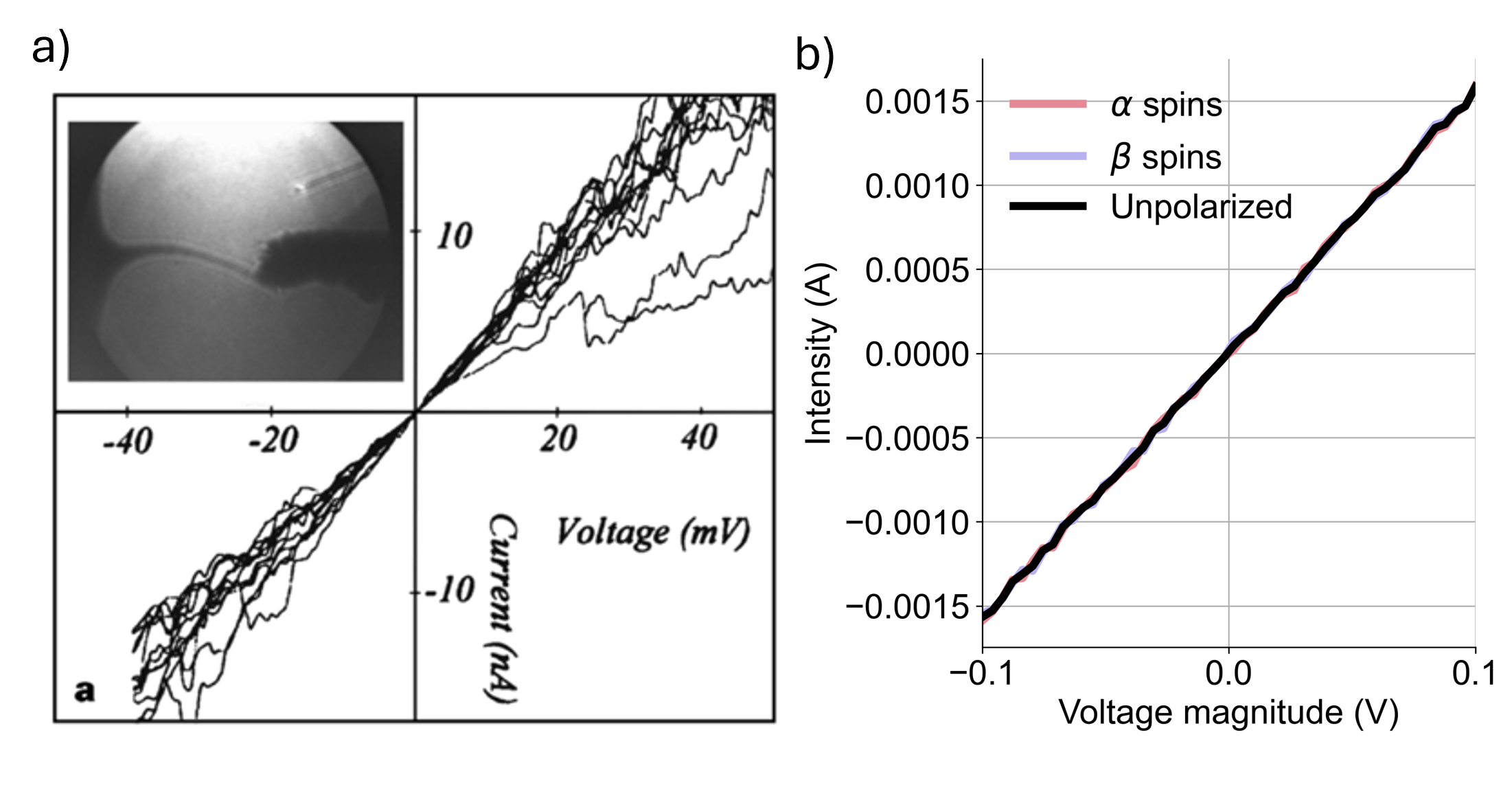}
    \caption{(a) Experimental $I$–$V$ response obtained for a 600-nm-long DNA rope. It is assumed an nearly-linear reponse at the low-bias region ($\pm20$~mV), from which a resistance of about $2.5$~M$\Omega$ is inferred.
    Reproduced with permission from \cite{Fink1999}
    (b) Current-voltage curves obtained from kinetic Monte Carlo simulations for the $\alpha$ (red), $\beta$ (purple), and unpolarized spin channels for an L-type helical molecule with $N=1000$ electrons, 44 positions, with both $\alpha$ and $\beta$ electrons starting at position 22, $D=0.2$, and ${\mathrm{emcha}}=0.1$ and time step length $\Delta(t)=10^{-17 }$ s. This small eMChA value results in a negligible spin-filtering effect. The simulation was performed for $N_{\mathrm{step}}=27000$ steps over 50 voltage values between $-0.1$ and $0.1$~V.}
    \label{fig:dna-current}
\end{figure*}

We illustrate the code developed so far by applying it to a DNA nanowire, as studied in the seminal experiment by Fink et al. \cite{Fink1999} They measured the current through a long, thin DNA nanowire, approximately 600~nm long and a few molecules wide, over a voltage range between $-50$~mV and $50$~mV. The measured response was linear, although noisy, with currents ranging between approximately $-15$~nA and $15$~nA, as shown in Figure \ref{fig:dna-current}(a).

For our simulation to reproduce representative experiments, we need to decide the values of a series of parameters. First, the number of molecules $N$, knowing that the total current will be directly proportional to $N$. Then, crucially, the spatial partition of our molecule, the number of distinct regions the spin can reside in. Fine partitions have a very significant calculation cost. This means that, except for the shortest molecules, this should usually be an effective number, unless one needs to extract some key insight that depends explicitly on the partition. This combination of assumptions results in the correct I-V response, but a number of the factors are either linearly or inversely linearly contributing to the current intensity. For example, the current is proportional to the number of electrons, i.e. 10 molecules with 1 electron per molecule produce the same current as 5 molecules with 2 electrons per molecule. Similarly, the diffusion parameter is only meaningful with respect to the time step length and the spatial partition: $D=0.5$ every $10^{-17}$~s results in the same current as $D=0.25$ every $0.5\cdot10^{-17}$~s (except for numerical errors). 

Therefore, the diffusion parameter and the characteristic hopping time can be adjusted to obtain the desired resistance for the simulated device. In the low-voltage and low-diffusion regime considered here, the current intensity displays the following dependencies:
\begin{itemize}
    \item linear dependence on the number of molecules, $I\propto N$;
    \item linear dependence on the diffusion parameter $D$, i.e. on the probability of an electron moving during the characteristic attempt time, $I\propto D$;
    \item inverse linear dependence on the characteristic attempt time, $I\propto 1/\tau$;
    \item inverse quadratic dependence on the number of spatial positions $n$ in the molecule, corresponding to the spatial partition, $I\propto 1/n^2$.
    \end{itemize}
These relations are qualitatively explained and numerically demonstrated in the Supplementary Information.

These scaling relations allow us to avoid the relatively large computational cost, and associated climate footprint, that would be required to simulate the current through a bundle of DNA molecules, each approximately 1800 base pairs long, and to resolve a relatively weak current of the order of nA. Instead, we can model a computationally inexpensive system that conducts much more intensely and subsequently obtain the appropriate experimental scale by applying the known scaling relations.

Of course, a smooth signal from a high-resistance system could, in principle, always be recovered by increasing the simulation time in proportion to the characteristic time step.  However, as the rate of hopping events decreases with increasing resistance, the noise eventually overwhelms the signal. The brute force approach of increasing the simulation time would provide the desired current resolution, whereas this gives the subjective satisfaction of seeing the right order of magnitude on the screen, this approach yields no additional physical insight and incurs a significant computational cost. We therefore do not pursue it. Instead, we initially do a coarse partition of the molecule to obtain a quick result and subsequently benefit from the known proportionality relations explained above to translate our result into the experimental case.

As a start, we employ the following coarse-grain model. We work with 1000 molecules divided into 44 number of spatial positions, with only one simultaneous electron per molecule, either $\alpha$ or $\beta$. Each electron has a probability of 20\% ($D=0.2$) of moving from one region to a neighbouring one after a time step of $10^{-17}$~s. With this parameter set, we obtain the behavior depicted in Figure \ref{fig:dna-current}(b). We can then translate these parameters, to the best of our knowledge, to the experimental system. We assume approximately 200 DNA molecules, corresponding to "a few" molecules as described by Fink et al. \cite{Fink1999}. If the form factor observed in their electron microscopy is approximately 20, the 600~nm DNA rope can be estimated to be approximately 30~nm wide, corresponding to an area of roughly $700$~nm$^2$ and approximately 200 DNA double strands. Each molecule is divided into 1800 base pairs, corresponding to a 600~nm long DNA helix in the most common B-form structure. As in the coarse-grained model, we consider only one simultaneous electron per molecule, either $\alpha$ or $\beta$, but now use a diffusion parameter of $D=0.01$, corresponding to a 1\% probability of an electron jump between base pairs per time step. The characteristic hopping time is left as a free parameter to fit the experimental current.

For the original time step of $10^{-17}$~s, this parameter set would give a current of $I=0.18\cdot10^{-8}$~A at $V=0.15$~V. To match the experimental current of $I\simeq10$ nA at 20 mV, this corresponds to a residence time of the order of $10^{-18}$~s in each base pair before reaching a probability of 1\% for an electron to jump to the neighboring base pair. The experimental system can thus be represented by a computationally inexpensive coarse-grained model, while the experimentally relevant current and resistance scales are recovered through the established parameter dependencies.

\subsection{Introducing spin}

In the simulation, the dynamical state of the system is represented by matrices that store the position of each electron along its molecular strand at every time step. In addition to the spatial degree of freedom, each electron is assigned a fixed spin label, allowing spin-resolved transport to be explicitly tracked throughout the simulation.

Electron transport is simulated independently for two spin channels, denoted here as $\alpha$ and $\beta$. These labels correspond to the two possible projections of the electron spin along a fixed quantization axis, with $\alpha$ spins defined as parallel and $\beta$ spins as antiparallel to that axis. The choice of quantization axis is fixed throughout the simulation and serves solely as a reference for defining spin-dependent transport channels; it does not imply the presence of a dynamically evolving magnetic field or coherent spin dynamics.

For each spin channel, hopping events that carry electrons across the left and right boundaries of the molecular strand are recorded separately. These boundary-crossing events provide an operational definition of spin-resolved electrical currents, defined as the net imbalance between crossings at the two ends over the duration of the simulation. Comparing the resulting currents for the two spin channels directly quantifies the spin selectivity of the transport process.

In the present implementation, the spin degree of freedom influences transport only through spin-dependent hopping probabilities, while spin-flip processes are neglected.The spin label of each electron is therefore conserved throughout the simulation. This corresponds to the assumption that spin relaxation occurs on time scales longer than those governing charge transport, or that its effects can be effectively absorbed into phenomenological transport parameters. Explicitly keeping track of the spin states nevertheless renders the framework readily extensible: stochastic spin dynamics, spin-flip events, or coupling to external fields could be incorporated in future versions of the model if required by the relevant physical regime.


\subsection{Recovering eMChA: symmetry difference with CISS}

Spin-dependent transport effects are introduced in the model by multiplying the energy difference governing the ratio of elementary hopping events in different directions, $\Delta_E$, with a term that is either larger or smaller than 1 depending on the combined action of spin, helicity, and transport direction. Specifically, the factor is obtained by an exponential where the sign factor of the exponent is taken as $(-1)^{s \cdot h \cdot v}$, where the spin variable $s$ is defined as $+1$ for electrons with $\alpha$ spins and $-1$ for electrons with $\beta$ spins, the helicity direction index $h$ is $+1$ for left-handed and $-1$ for right-handed molecular chiral structures, and the velocity index $v$ distinguishes motion parallel ($+1$) or antiparallel ($-1$) to the chosen spin quantization axis. This construction captures the symmetry requirements expected for magnetochiral transport phenomena, where the transport response changes sign under reversal of any one of these three quantities.

To model electric magnetochiral anisotropy, we assume that the relevant spin-dependent contribution originates from spin-orbit coupling to the magnetic field generated by charge motion along a chiral trajectory. Importantly, an electron undergoing mostly random, back-and-forth hopping over short distances does not generate a net magnetic field that can couple to its own spin: a linear charge movement generates a cylindrical magnetic field that vanishes at its axis. This situation corresponds to the low-voltage, diffusion-dominated regime of transport, where the electric current is negligible and the eMChA signal is therefore also expected to vanish. This behavior is consistent with experimental observations showing the effective suppression of electrical magnetochiral effects in the linear-response limit of the current.

In contrast, when transport is driven by a substantial applied voltage, electrons acquire a net directional motion along the molecular helix: while there will be randomness, there is a marked preferential direction for the motion. In this regime, the system behaves analogously to a current along a coil, where the circulating charge density on a part of the coil generates a magnetic field that is felt on a different part of the coil. Thus, it can act on the same electronic states, enabling a coupling between orbital motion and spin. One possible way to model this effect would be to explicitly track the trajectory history of each electron and activate the spin-orbit coupling term only after a threshold number of consecutive hops in the same direction, mimicking the buildup of a persistent current. This approach has been conceptuallized in the Hamiltonian proposed by Fransson, where spin-orbit coupling connects next-nearest-neighbor sites, while nearest-neighbor hopping remains spin independent.\cite{fransson2024current}

However, explicitly storing and processing trajectory histories would significantly increase the computational complexity of the simulation. Instead, we adopt a more compact phenomenological approach that exploits a key property of the present model: the net electrical current generated by an applied voltage scales as a hyperbolic tangent function of the voltage. Since the magnetic field associated with the orbital motion is itself proportional to the current, we model the effective spin-orbit contribution to the magnetochiral energy term as also being modulated by a $\tanh(V)$ dependence. This choice ensures that the eMChA contribution vanishes at zero voltage, grows linearly at small voltages where the current is proportional to the electric field, and saturates at large enough voltages, when the current reaches its maximum value.

\begin{figure}
    \centering
    \includegraphics[width=\linewidth]{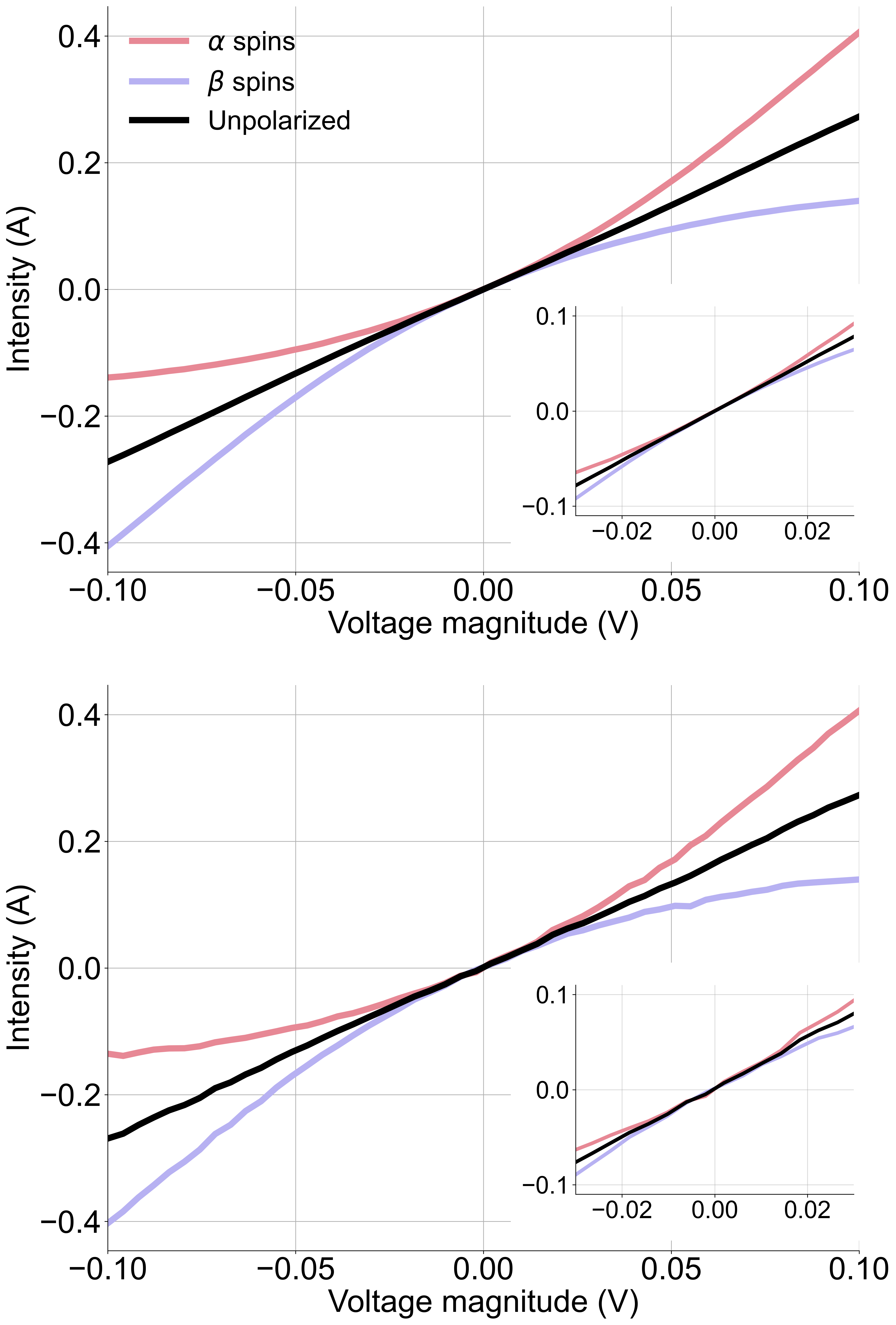}
    \caption{Current-voltage curve obtained from the kinetic Monte Carlo simulation for $\alpha$ (red curve) and $\beta$ (purple curve) spin channels, together with the corresponding unpolarized current. The results are shown for a system comprising $N=1000$ electrons moving through identical molecular strands for $N_{step}=50000$ steps (top panel) or $N_{step}=500$ steps (bottom panel), in an L-type helical molecule with 5 possible positions and a dimensionless diffusion coefficient $D=0.5$ controlling the overall hopping rate, and a phenomenological magnetochiral anisotropy effect ${\mathrm{emcha}}=30$ that introduces spin-dependent asymmetry in the hopping probabilities and for a total of 50 different voltage values between -0.1 and 0.1 V. The currents are reported in arbitrary units and exhibit a nonlinear voltage dependence with spin-resolved asymmetry characteristic of electric magnetochiral anisotropy. The insets contain zooms at low voltage. }
    \label{fig:emcha}
\end{figure}

The resulting behavior of the model reproduces the key qualitative features expected for electric magnetochiral anisotropy, as illustrated in Fig.~\ref{fig:emcha}. At low voltages, electrons with $\alpha$ and $\beta$ spin projections respond almost identically to the applied bias, resulting in an almost linear current-voltage response and a vanishing eMChA signal, as seen more clearly in the insets. At higher voltages, the spin-dependent asymmetry becomes apparent: for the chosen (L) molecular helicity, $\alpha$ electrons exhibit a lower resistance at positive bias and a higher resistance at negative bias, while $\beta$ electrons show the opposite behavior. The figure also illustrate the stochastical nature of our calculations: if we choose to simulate a mere 500 time steps rather than 50000, one can already see the appearance of some noise. The calculation takes seconds or a few minutes, in any case.

An open question remains, however, as to why CISS experiments do not display the same low-voltage behavior, instead presenting an essentially linear spin-filtering response. This discrepancy will hopefully be addressed in future versions of the simulator. At present, we can conclude that the model provides a reasonable phenomenological description of a mechanism that naturally produces eMChA without explicitly including the electrodes. In agreement with the symmetry considerations discussed by Rikken, \cite{Rikken2023} the resulting transport displays the characteristic eMChA behavior shown in Figure~\ref{fig:Rikken}.

\subsection{CISS vs. eMChA}
A final remark concerns the role of the external magnetic field. In the present simulator, an explicit magnetic-field dependence does not need to be introduced, as the expected linear behavior emerges naturally from the underlying assumptions. This can be understood by considering two limiting regimes: the regime where one is at near-saturation magnetization limit, as is assumed in CISS experiments, and the regime with a near-linear magnetic response, as happens in eMChA experiments. In paramagnets, these would correspond to the high-field, low-temperature (HFLT) limit and the low-field, high-temperature (LFHT) limit, respectively.

In the HFLT limit, the Zeeman energy dominates over thermal fluctuations, and the electronic spins become fully aligned with the external magnetic field. This results in a near-saturation magnetization, which is also found, in absence of external magnetic field, in permanent magnets. This is the case of CISS experiments, typically conducted at room temperature but obtaining a highly spin-polarized currents which are typically generated using ferromagnetic or otherwise spin-selective electrodes rather than by applying an explicit external magnetic field or by lowering the temperature.

By contrast, eMChA experiments are commonly performed in the LFHT regime, where the Zeeman energy remains much smaller than the thermal energy and the magnetic response is linear. In this limit, the net spin polarization is weak and varies linearly with the applied magnetic field. When transport occurs through a chiral conductor that exhibits different resistances for opposite spin orientations, the contributions from the two spin channels largely cancel for weakly polarized currents. The residual signal is therefore proportional to the small spin imbalance.

Since this spin excess itself scales linearly with the magnetic field in the low-field regime, the resulting electric magnetochiral anisotropy signal also exhibits a linear dependence on $B$. This argument explains why an explicit magnetic-field term is not required in the simulation to recover the experimentally observed magnetic-field scaling of the eMChA effect. As we will see in the next section, we recover the magnetic field dependence simply by calculating the global current as a weighted average of the $\alpha$ and $\beta$ currents for a given magnetic field (i.e. Zeeman energy) and temperature.

\subsection{Quantifying eMChA: the interaction parameters $\gamma$ and $\Delta R$}

Once we obtain the electric current for $\alpha$ and $\beta$ spins at any given applied voltage, it is almost trivial to estimate the relative resistance anisotropy $\Delta R$ as defined by Pop et al. as the change in resistance measured when switching the applied voltage to obtain a change in the sign of the current intensity, at a given applied magnetic field:\cite{Pop2014}
\begin{equation}
    \Delta R(I,B) \equiv R(I,B) - R(-I,B)
    \label{DeltaR_eMChA}
\end{equation}

From $\Delta_R$, Pop et al. indirectly define an eMChA anisotropy parameters $\gamma^{D/L}$ for right- and left-handed molecules as:\cite{Pop2014}
\begin{equation}
    \frac{\Delta R}{R} = 2\gamma^{D/L} \cdot B \cdot I
    \label{DeltaR_eMChA}
\end{equation}

Experimentally, $\gamma^{D/L}$ can be determined as follows: one switches the sign both of the magnetic field and of the applied voltage, and then corrects the voltage until one obtains also an exact sign change for the current intensity. $\gamma^{D/L}$ is then given by:\cite{Pop2014}
\begin{equation}
    \gamma^{D/L} = \frac{V^{D/L}_\mathrm{eMChA}}{2GRBI^2}
    \label{DeltaR_eMChA}
\end{equation}
where $V_\mathrm{eMChA}$ is the voltage correction beyond a voltage sign switch needed to obtain a sign switch in the $I'=-I$ after switching the orientation of the external magnetic field. $G$ is an electronic preamplifier gain in the circuit employed in the experiment we employ as an example to illustrate the calculation,\cite{Pop2014} where it takes the value $G=185$.

At this point one simply needs to compute the total current in different conditions, in particular for two values of the intensity, $I$ and $I'=-I$. For a given magnetic field $B$, and assuming metallic electrodes presenting conduction electrons whose magnetic moment we estimate as a Landé's $g=2$ on a spin $S=1/2$, it is immediate to compute the Zeeman energy difference between alpha and beta spins, as:
\begin{equation}
    E_\mathrm{Zeeman} = 2\cdot S\cdot g \cdot \mu_B\cdot B = 2\cdot\mu_B \cdot B
\end{equation}
where $\mu_B$ is Bohr's magneton.

From $E_\mathrm{Zeeman}$ and room temperature $T=300$~K, we estimate Boltzmann's population distribution for $\alpha$ and $\beta$ spins. Given that we already showed how to estimate the electric current for the different spin states, this is all we need to obtain the total electric current, by applying a weighted average. In turn, repeating the process for $I'=-I$ results in $\Delta R$ per equation \ref{DeltaR_eMChA}.

\begin{figure}
    \centering
    \includegraphics[width=\linewidth]{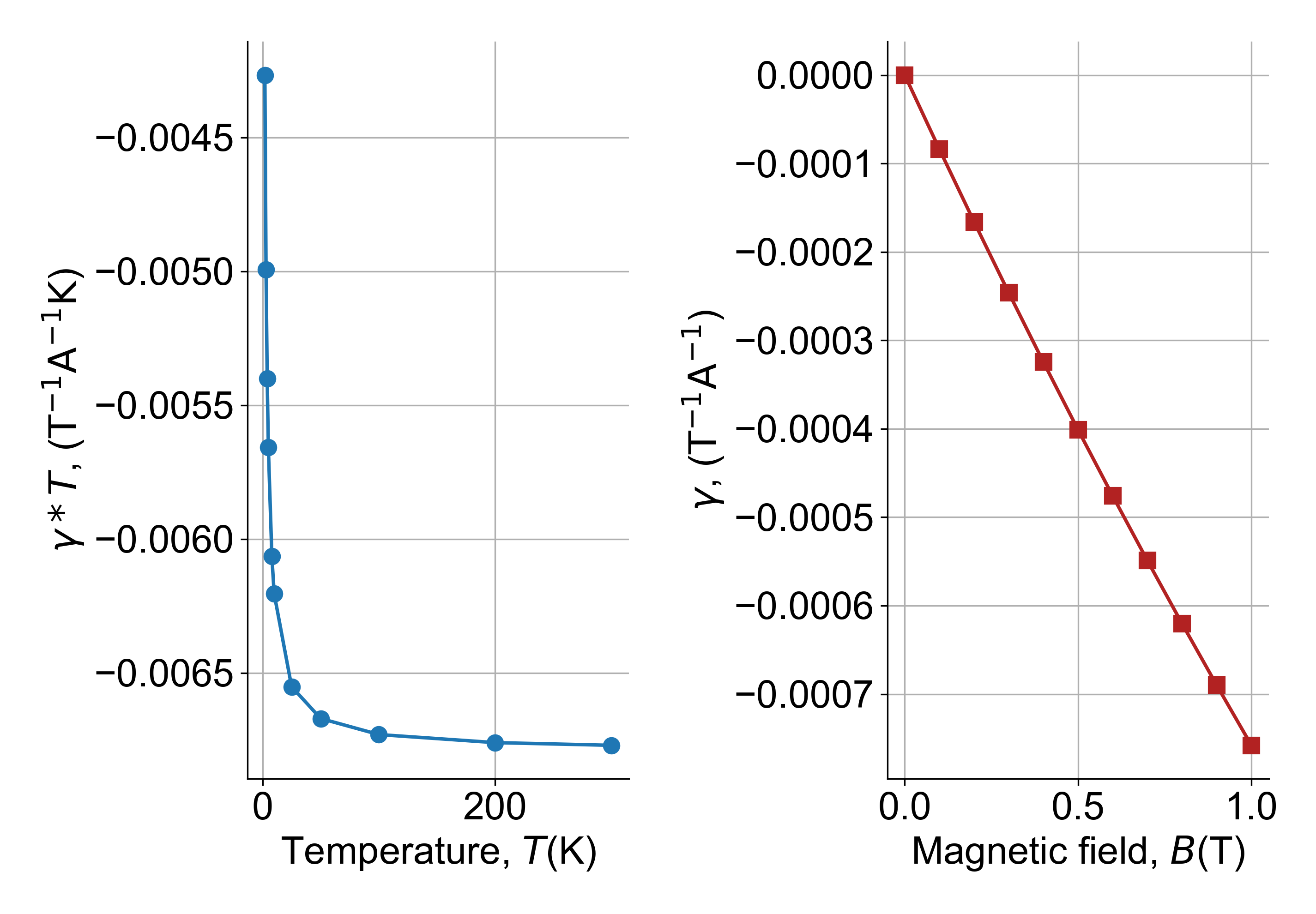}
    \caption{$\gamma$ as a function of $T$ between 1 and 300 K, at $B=1$T (left) and as a function of $B$ between 0 and 1 T, at $T=8$K (right). On the left panel $\gamma\cdot T$ is plotted rather than $\gamma$ to visually evidence the behavior $\gamma=C/T$ at high $T$ and the deviation at low $T$.}
    \label{fig:gammavsTB}
\end{figure}

As we see in figure \ref{fig:gammavsTB}, $\gamma$ is directly proportional to the magnetic field $B$, for low enough values of $B$ and inversely proportional to the temperature $T$ for high enough values of $T$. This is nothing more than a manifestation of the exponent in Boltzmann's formula $\Delta E / (kT)$, in the regime of small population differences i.e. small $\Delta E / (kT)$. Spins $\alpha$ and $\beta$ present distinct resistances when moving through a chiral medium, but this phenomenon can only manifest experimentally if the populations of $\alpha$ and $\beta$ are different, in eMChA experiments (or if we have a mechanism to effectively filter them, in CISS experiments). Indeed, the phenomenon is directly proportional to this population difference or initial spin polarization, and this is the effect that our simulations recover.

\section{Conclusions}

We developed a simple model with an inexpensive implementation, which naïvely could be seen as a microscopic implementation of either eMChA or CISS. From its simple set of rules we find that the behavior that emerges corresponds to eMChA, therefore shining some light on this phenomenon, and evidencing that something else is needed to recover the experimental observations of CISS experiments.

The current version of the model presented here covers only a basic functionality compared with its potential. Nevertheless, it already reproduces both the independent behavior of conventional electrical resistors and and the qualitative dependence of eMChA on voltage and magnetic field. These results provide a useful test of the underlying stochastic framework: by introducing spin and chirality with a minimal set of assumptions, the model naturally recovers the characteristic symmetry and nonlinear transport behavior associated with eMChA. At the same time, the absence of the characteristic low-voltage spin filtering observed in CISS experiments as well as the mismatch in filtering symmetry highlights fundamental differences between the two phenomena, and points to an additional complexity behind the CISS effect.

The model is inherently expandable, allowing such extensions to be developed while preserving its frugal character. This is supported by our previous kMC code DAISY, \cite{GutierrezFinol2025DAISY} which is able to reproduce experimental ac phenomena, including the in-phase and out-of-phase susceptibility of single-molecule magnets. Similarly, the related CupFlow code can reproduce experiments involving alternating increasing and decreasing voltage, where charge accumulation rather than charge flow is measured. \cite{GutierrezFinol2025CupFlow} The three approaches could therefore be combined to explore other members of the magnetochiral family, including dielectric magnetochiral anisotropy, the more recently investigated extension of magnetochiral phenomena. \cite{rikken2022dielectric}

More broadly, the results suggest that a minimal stochastic transport framework can provide a useful platform for testing different mechanisms of chiral spin-dependent transport. In this sense, the present model is not intended to provide a complete microscopic description of CISS or eMChA, but rather to establish which experimentally observed features can emerge from a simple set of transport rules and which require additional physics. This distinction provides a practical starting point for future extensions of the model.

\section*{Theoretical framework and computational methods}

\subsection*{Current, voltage and saturation}

Before considering any interaction with chirality, a system with charged carriers under an applied voltage \(V\) exhibits an intrinsic asymmetry between forward (rightward) and backward (leftward) transition probabilities. Assuming a Boltzmann distribution for the transition rates, the probability for an electron to move in the direction of the electric field (\(P_{+}\)) or against it (\(P_{-}\)) can be written as:
\begin{equation}
P_{+} = \frac{e^{-\beta qV}}{e^{-\beta qV} + e^{\beta qV}}, \qquad
P_{-} = \frac{e^{\beta qV}}{e^{-\beta qV} + e^{\beta qV}},
\end{equation}
where \(\beta = 1/(k_B T)\) is the inverse thermal energy, \(q\) is the electron charge, and \(T\) is the temperature. \\
The net directional imbalance, and therefore the net current, is proportional to the difference in these probabilities:
\begin{equation}
I \propto |\Delta P| = |P_{+} - P_{-}| = \tanh\left( \frac{qV}{k_B T} \right).
\end{equation}
This result shows a hyperbolic tangent profile, with the probability imbalance increasing linearly with voltage at low bias and saturating at high bias due to thermal broadening. 

\subsection*{Chirality and spin-orbit-coupling}

Based on this outcome, the emergence of spin-dependent transport in chiral systems can be qualitatively understood as a two-step process.
Firstly, an applied voltage \(V\) induces a net current due to the asymmetry between forward and backward transition probabilities. Secondly, this current generates a magnetic field via Ampère's law.

This magnetic field points perpendicular to the direction of charge transport and to the chiral axis of the molecule. As a result, electrons moving through the chiral medium experience an effective magnetic field whose magnitude is proportional to the current, and hence to \(\tanh(qV/k_BT)\). This magnetic field couples to the electron spin via spin-orbit coupling (SOC), producing a spin-dependent energy correction.\\
Therefore, the total energy variation associated with a transition should include not only the purely electric contribution, \(qV\), but also a spin-dependent term of the form \(\mu \cdot B\), where \(\mu\) is the electron magnetic moment and \(B\) is the effective magnetic field generated by the current. Phenomenologically, we model this additional contribution $c_{\text{eMChA}}$ as:
\begin{equation}
c_{\text{eMChA}} = \Gamma \cdot s \cdot \tanh\left( \frac{qV}{k_B T} \right),
\label{eq:gammastanh}
\end{equation}
where \(\Gamma\) is a parameter capturing the maximum magnitude of the eMChA-induced spin-orbit energy shift and \(s = \pm 1\) denotes the electron spin orientation. This contribution represents the effective spin-dependent potential arising from the SOC mediated by the current-induced magnetic field. 

Thus, the total energy associated with a transition of an electron with spin \(s\) in direction \(d = \pm 1\) (forward or backward) is then given by:
\begin{equation}
\Delta E = qV \cdot e^{c_{\text{eMChA}}}
\end{equation}
meaning eMChA can either intensify or debilitate the energy bias between two adjacent sites that is originally created by the external voltage.

We can see that three signs are multiplied in equation \ref{eq:gammastanh}, exactly as in the simplified expression $(-1)^{s·c·v}$ introduced above, electron spin $s$ times molecular chirality $c$ times electron velocity $v$. Indeed it is a different expression for the same physics, since $c$ is proportional to $\Gamma$ and $v$ is proportional to $\tanh\left( \frac{qV}{k_B T} \right)$. The product between these three parities determines whether a given current direction is favoured or disfavoured for a given electron spin in a given molecular chirality.

\section*{Acknowledgments}

The authors gratefully acknowledge funding by the Generalitat Valenciana (GVA) CIDEGENT/2021/018 grant. S.G.S. acknowledges funding from Taighde Éireann – Research Ireland under grant No. GOIPD/2024/676, and the European Union’s Horizon 2024 research and innovation programme under the Marie Skłodowska-Curie grant No. 101209627. We thank G. L. Rikken for his insightful comments and fruitful discussions.

\section*{Author declarations}

\subsection*{Conflicts of interest}
The authors declare no relevant conflict of interest about the present work.

\subsection*{Author contributions}
Conceptualization: AGA, SGS
; Data curation: SGS
; Formal analysis: SGS, AGA
; Funding acquisition: SGS, AGA
; Investigation: SGS, AGA
; Methodology: SGS, GGF, AMM
; Project administration: SGS, AGA 
; Software: AMM, SGS, GGF
; Resources: SGS
; Supervision: SGS, AGA 
; Validation: SGS, AGA
; Visualization: AMM, GGF, SGS
; Writing – original draft: GGF, AGA, AMM, SGS. 
; Writing – review \& editing: AGA, SGS.



\section*{Data Availability}

All simulation and analysis scripts used in this work are openly accessible in the 
\href{https://github.com/silsgs/kMC4eMCHA.git}{kMC4eMCHA} GitHub repository. 
The repository includes the full simulation framework, parameter files, and example scripts.

\section*{References}

\bibliography{kMC4CISS}

@article{Wolf2001,
  title = {Spintronics: A Spin-Based Electronics Vision for the Future},
  volume = {294},
  ISSN = {1095-9203},
  url = {http://dx.doi.org/10.1126/science.1065389},
  DOI = {10.1126/science.1065389},
  number = {5546},
  journal = {Science},
  publisher = {American Association for the Advancement of Science (AAAS)},
  author = {Wolf,  S. A. and Awschalom,  D. D. and Buhrman,  R. A. and Daughton,  J. M. and von Moln\'ar,  S. and Roukes,  M. L. and Chtchelkanova,  A. Y. and Treger,  D. M.},
  year = {2001},
  month = nov,
  pages = {1488–1495}
}

@article{Zutic2004,
  title = {Spintronics: Fundamentals and applications},
  volume = {76},
  ISSN = {1539-0756},
  url = {http://dx.doi.org/10.1103/RevModPhys.76.323},
  DOI = {10.1103/revmodphys.76.323},
  number = {2},
  journal = {Reviews of Modern Physics},
  publisher = {American Physical Society (APS)},
  author = {\u{Z}uti\'c,  Igor and Fabian,  Jaroslav and Das Sarma,  S.},
  year = {2004},
  month = apr,
  pages = {323–410}
}

@article{GutierrezFinol2025CupFlow,
    author = {Gutiérrez-Finol, Gerliz M. and Zinovjev, Kirill and Gaita-Ariño, Alejandro and Cardona-Serra, Salvador},
    title = {A scalable kinetic Monte Carlo platform for charge transport dynamics in polymer-based memristive systems},
    journal = {Materials Chemistry Frontiers},
    volume = {10},
    number = {9},
    pages = {1437-1445},
    year = {2026},
    month = {05},
    issn = {2052-1537},
    doi = {10.1039/d5qm00811e},
    url = {https://doi.org/10.1039/d5qm00811e},
    eprint = {https://pubs.rsc.org/qm/article-pdf/10/9/1437/12764986/d5qm00811e.pdf},
}

@article{Hirohata2020,
  title = {Review on spintronics: Principles and device applications},
  volume = {509},
  ISSN = {0304-8853},
  url = {http://dx.doi.org/10.1016/j.jmmm.2020.166711},
  DOI = {10.1016/j.jmmm.2020.166711},
  journal = {Journal of Magnetism and Magnetic Materials},
  publisher = {Elsevier BV},
  author = {Hirohata,  Atsufumi and Yamada,  Keisuke and Nakatani,  Yoshinobu and Prejbeanu,  Ioan-Lucian and Diény,  Bernard and Pirro,  Philipp and Hillebrands,  Burkard},
  year = {2020},
  month = sep,
  pages = {166711}
}

@article{Manchon2019,
  title = {Current-induced spin-orbit torques in ferromagnetic and antiferromagnetic systems},
  volume = {91},
  ISSN = {1539-0756},
  url = {http://dx.doi.org/10.1103/RevModPhys.91.035004},
  DOI = {10.1103/revmodphys.91.035004},
  number = {3},
  journal = {Reviews of Modern Physics},
  publisher = {American Physical Society (APS)},
  author = {Manchon,  A. and Železný,  J. and Miron,  I. M. and Jungwirth,  T. and Sinova,  J. and Thiaville,  A. and Garello,  K. and Gambardella,  P.},
  year = {2019},
  month = sep 
}

@article{Naaman2015,
  title = {Spintronics and Chirality: Spin Selectivity in Electron Transport Through Chiral Molecules},
  volume = {66},
  ISSN = {1545-1593},
  url = {http://dx.doi.org/10.1146/annurev-physchem-040214-121554},
  DOI = {10.1146/annurev-physchem-040214-121554},
  number = {1},
  journal = {Annual Review of Physical Chemistry},
  publisher = {Annual Reviews},
  author = {Naaman,  Ron and Waldeck,  David H.},
  year = {2015},
  month = apr,
  pages = {263–281}
}

@article{Gupta2024,
  title = {The chirality-induced spin selectivity effect in asymmetric spin transport: from solution to device applications},
  volume = {15},
  ISSN = {2041-6539},
  url = {http://dx.doi.org/10.1039/D4SC05736H},
  DOI = {10.1039/d4sc05736h},
  number = {45},
  journal = {Chemical Science},
  publisher = {Royal Society of Chemistry (RSC)},
  author = {Gupta,  Ritu and Balo,  Anujit and Garg,  Rabia and Mondal,  Amit Kumar and Ghosh,  Koyel Banerjee and Chandra Mondal,  Prakash},
  year = {2024},
  pages = {18751–18771}
}

@article{Bloom2024,
  title = {Chiral Induced Spin Selectivity},
  volume = {124},
  ISSN = {1520-6890},
  url = {http://dx.doi.org/10.1021/acs.chemrev.3c00661},
  DOI = {10.1021/acs.chemrev.3c00661},
  number = {4},
  journal = {Chemical Reviews},
  publisher = {American Chemical Society (ACS)},
  author = {Bloom,  Brian P. and Paltiel,  Yossi and Naaman,  Ron and Waldeck,  David H.},
  year = {2024},
  month = feb,
  pages = {1950–1991}
}

@article{Naaman2012,
  title = {Chiral-Induced Spin Selectivity Effect},
  volume = {3},
  ISSN = {1948-7185},
  url = {http://dx.doi.org/10.1021/jz300793y},
  DOI = {10.1021/jz300793y},
  number = {16},
  journal = {The Journal of Physical Chemistry Letters},
  publisher = {American Chemical Society (ACS)},
  author = {Naaman,  R. and Waldeck,  David H.},
  year = {2012},
  month = jul,
  pages = {2178–2187}
}

@Article{Naaman2016,
author ="Michaeli, Karen and Kantor-Uriel, Nirit and Naaman, Ron and Waldeck, David H.",
title  ="The electron{'}s spin and molecular chirality – how are they related and how do they affect life processes?",
journal  ="Chem. Soc. Rev.",
year  ="2016",
volume  ="45",
issue  ="23",
pages  ="6478-6487",
publisher  ="The Royal Society of Chemistry",
doi  ="10.1039/C6CS00369A",
url  ="http://dx.doi.org/10.1039/C6CS00369A"
}

@article{Naaman2019,
  title = {Chiral molecules and the electron spin},
  volume = {3},
  ISSN = {2397-3358},
  url = {http://dx.doi.org/10.1038/s41570-019-0087-1},
  DOI = {10.1038/s41570-019-0087-1},
  number = {4},
  journal = {Nature Reviews Chemistry},
  publisher = {Springer Science and Business Media LLC},
  author = {Naaman,  Ron and Paltiel,  Yossi and Waldeck,  David H.},
  year = {2019},
  month = mar,
  pages = {250–260}
}

@article{Rikken1997,
  title = {Observation of magneto-chiral dichroism},
  volume = {390},
  ISSN = {1476-4687},
  number = {6659},
  journal = {Nature},
  publisher = {Springer Science and Business Media LLC},
  author = {Rikken,  G. L. J. A. and Raupach,  E.},
  year = {1997},
  month = dec,
  pages = {493–494}
}

@article{Pop2014,
  title = {Electrical magnetochiral anisotropy in a bulk chiral molecular conductor},
  volume = {5},
  ISSN = {2041-1723},
  url = {http://dx.doi.org/10.1038/ncomms4757},
  DOI = {10.1038/ncomms4757},
  number = {1},
  journal = {Nature Communications},
  publisher = {Springer Science and Business Media LLC},
  author = {Pop,  Flavia and Auban-Senzier,  Pascale and Canadell,  Enric and Rikken,  Geert L. J. A. and Avarvari,  Narcis},
  year = {2014},
  month = may 
}

@article{Rikken2001,
  title = {Electrical Magnetochiral Anisotropy},
  volume = {87},
  ISSN = {1079-7114},
  url = {http://dx.doi.org/10.1103/PhysRevLett.87.236602},
  DOI = {10.1103/physrevlett.87.236602},
  number = {23},
  journal = {Physical Review Letters},
  publisher = {American Physical Society (APS)},
  author = {Rikken,  G. L. J. A. and F\"{o}lling,  J. and Wyder,  P.},
  year = {2001},
  month = nov 
}

@Article{Rikken2023,
author={Rikken, G. L. J. A.
and Avarvari, N.},
title={Comparing Electrical Magnetochiral Anisotropy and Chirality-Induced Spin Selectivity},
journal={The Journal of Physical Chemistry Letters},
year={2023},
month={Nov},
day={02},
publisher={American Chemical Society},
volume={14},
number={43},
pages={9727-9731},
doi={10.1021/acs.jpclett.3c02546},
url={https://doi.org/10.1021/acs.jpclett.3c02546}
}

@article{Giaconi2024,
  title = {Spin polarized current in chiral organic radical monolayers},
  volume = {12},
  ISSN = {2050-7534},
  url = {http://dx.doi.org/10.1039/D4TC00944D},
  DOI = {10.1039/d4tc00944d},
  number = {27},
  journal = {Journal of Materials Chemistry C},
  publisher = {Royal Society of Chemistry (RSC)},
  author = {Giaconi,  Niccolò and Lupi,  Michela and Das,  Tapan Kumar and Kumar,  Anil and Poggini,  Lorenzo and Viglianisi,  Caterina and Sorace,  Lorenzo and Menichetti,  Stefano and Naaman,  Ron and Sessoli,  Roberta and Mannini,  Matteo},
  year = {2024},
  pages = {10029–10035}
}

@article{Atzori2021,
author = {Atzori, Matteo and Train, Cyrille and Hillard, Elizabeth A. and Avarvari, Narcis and Rikken, Geert L. J. A.},
title = {Magneto-chiral anisotropy: From fundamentals to perspectives},
journal = {Chirality},
volume = {33},
number = {12},
pages = {844-857},
year = {2021}
}

@article{GutierrezFinol2026CupFlow,
  title = {A scalable kinetic Monte Carlo platform for charge transport dynamics in polymer-based memristive systems},
  ISSN = {2052-1537},
  url = {http://dx.doi.org/10.1039/D5QM00811E},
  DOI = {10.1039/d5qm00811e},
  journal = {Materials Chemistry Frontiers},
  publisher = {Royal Society of Chemistry (RSC)},
  author = {Gutiérrez-Finol,  Gerliz M. and Zinovjev,  Kirill and Gaita-Ariño,  Alejandro and Cardona-Serra,  Salvador},
  year = {2026}
}

@article{Ghazaryan2020,
  title = {Analytic Model of Chiral-Induced Spin Selectivity},
  volume = {124},
  ISSN = {1932-7455},
  url = {http://dx.doi.org/10.1021/acs.jpcc.0c02584},
  DOI = {10.1021/acs.jpcc.0c02584},
  number = {21},
  journal = {The Journal of Physical Chemistry C},
  publisher = {American Chemical Society (ACS)},
  author = {Ghazaryan,  Areg and Paltiel,  Yossi and Lemeshko,  Mikhail},
  year = {2020},
  month = may,
  pages = {11716–11721}
}

@article{Savi2025,
  title = {Chirality-Induced Spin Selectivity: A Minimal Model},
  volume = {16},
  ISSN = {1948-7185},
  url = {http://dx.doi.org/10.1021/acs.jpclett.5c01813},
  DOI = {10.1021/acs.jpclett.5c01813},
  number = {35},
  journal = {The Journal of Physical Chemistry Letters},
  publisher = {American Chemical Society (ACS)},
  author = {Savi,  Lorenzo and Celada,  Leonardo and Phan Huu,  D.K. Andrea and Chiesa,  Alessandro and Carretta,  Stefano and Painelli,  Anna},
  year = {2025},
  month = aug,
  pages = {9107–9115}
}

@article{Chiesa2025,
  title = {Chirality-Induced Spin Selectivity at the Molecular Level: A Different Perspective to Understand and Exploit the Phenomenon},
  volume = {16},
  ISSN = {1948-7185},
  url = {http://dx.doi.org/10.1021/acs.jpclett.5c00755},
  DOI = {10.1021/acs.jpclett.5c00755},
  number = {21},
  journal = {The Journal of Physical Chemistry Letters},
  publisher = {American Chemical Society (ACS)},
  author = {Chiesa,  Alessandro and Privitera,  Alberto and Garlatti,  Elena and Allodi,  Giuseppe and Bittl,  Robert and Wasielewski,  Michael R. and Sessoli,  Roberta and Carretta,  Stefano},
  year = {2025},
  month = may,
  pages = {5358–5372}
}

@article{Fransson2025,
  title = {Should it really be that hard to model the chirality induced spin selectivity effect?},
  volume = {1},
  ISSN = {3066-0017},
  url = {http://dx.doi.org/10.1063/5.0289548},
  DOI = {10.1063/5.0289548},
  number = {2},
  journal = {APL Computational Physics},
  publisher = {AIP Publishing},
  author = {Fransson,  Jonas},
  year = {2025},
  month = oct 
}

@article{Xu2023,
  title = {Chiral-induced spin selectivity in biomolecules,  hybrid organic–inorganic perovskites and inorganic materials: a comprehensive review on recent progress},
  volume = {10},
  ISSN = {2051-6355},
  url = {http://dx.doi.org/10.1039/D3MH00024A},
  DOI = {10.1039/d3mh00024a},
  number = {6},
  journal = {Materials Horizons},
  publisher = {Royal Society of Chemistry (RSC)},
  author = {Xu,  Yingdan and Mi,  Wenbo},
  year = {2023},
  pages = {1924–1955}
}

@article{Eckvahl2023,
  title = {Direct observation of chirality-induced spin selectivity in electron donor–acceptor molecules},
  volume = {382},
  ISSN = {1095-9203},
  url = {http://dx.doi.org/10.1126/science.adj5328},
  DOI = {10.1126/science.adj5328},
  number = {6667},
  journal = {Science},
  publisher = {American Association for the Advancement of Science (AAAS)},
  author = {Eckvahl,  Hannah J. and Tcyrulnikov,  Nikolai A. and Chiesa,  Alessandro and Bradley,  Jillian M. and Young,  Ryan M. and Carretta,  Stefano and Krzyaniak,  Matthew D. and Wasielewski,  Michael R.},
  year = {2023},
  month = oct,
  pages = {197–201}
}

@inbook{Camsari2022,
  title = {The Nonequilibrium Green Function (NEGF) Method},
  ISBN = {9783030798277},
  ISSN = {2522-8706},
  url = {http://dx.doi.org/10.1007/978-3-030-79827-7_44},
  DOI = {10.1007/978-3-030-79827-7_44},
  booktitle = {Springer Handbook of Semiconductor Devices},
  publisher = {Springer International Publishing},
  author = {Camsari,  Kerem Y. and Chowdhury,  Shuvro and Datta,  Supriyo},
  year = {2022},
  month = nov,
  pages = {1583–1599}
}

@article{FormentAliaga2022,
    author = {Forment-Aliaga, Alicia and Gaita-Ariño, Alejandro},
    title = "{Chiral, magnetic, molecule-based materials: A chemical path toward spintronics and quantum nanodevices}",
    journal = {Journal of Applied Physics},
    volume = {132},
    number = {18},
    pages = {180901},
    year = {2022},
    month = {11},
    issn = {0021-8979}
}

@article{Long2020,
  title = {Chiral-perovskite optoelectronics},
  volume = {5},
  ISSN = {2058-8437},
  number = {6},
  journal = {Nature Reviews Materials},
  publisher = {Springer Science and Business Media LLC},
  author = {Long,  Guankui and Sabatini,  Randy and Saidaminov,  Makhsud I. and Lakhwani,  Girish and Rasmita,  Abdullah and Liu,  Xiaogang and Sargent,  Edward H. and Gao,  Weibo},
  year = {2020},
  month = mar,
  pages = {423–439}
}

@article{TorresCavanillas2020,
author = {Torres-Cavanillas, Ramón and Escorcia-Ariza, Garin and Brotons-Alcázar, Isaac and Sanchis-Gual, Roger and Mondal, Prakash Chandra and Rosaleny, Lorena E. and Giménez-Santamarina, Silvia and Sessolo, Michele and Galbiati, Marta and Tatay, Sergio and Gaita-Ariño, Alejandro and Forment-Aliaga, Alicia and Cardona-Serra, Salvador},
title = {Reinforced room-temperature spin filtering in chiral paramagnetic metallopeptides},
journal = {Journal of the American Chemical Society},
volume = {142},
number = {41},
pages = {17572},
year = {2020}
}

@article{HuiziRayo2020,
  title = {An Ideal Spin Filter: Long-Range,  High-Spin Selectivity in Chiral Helicoidal 3-Dimensional Metal Organic Frameworks},
  volume = {20},
  ISSN = {1530-6992},
  url = {http://dx.doi.org/10.1021/acs.nanolett.0c02349},
  DOI = {10.1021/acs.nanolett.0c02349},
  number = {12},
  journal = {Nano Letters},
  publisher = {American Chemical Society (ACS)},
  author = {Huizi-Rayo,  Uxua and Gutierrez,  Junkal and Seco,  Jose Manuel and Mujica,  Vladimiro and Diez-Perez,  Ismael and Ugalde,  Jesus M. and Tercjak,  Agnieszka and Cepeda,  Javier and San Sebastian,  Eider},
  year = {2020},
  month = nov,
  pages = {8476–8482}
}

@article{Gohler2011,
  title = {Spin Selectivity in Electron Transmission Through Self-Assembled Monolayers of Double-Stranded DNA},
  volume = {331},
  ISSN = {1095-9203},
  url = {http://dx.doi.org/10.1126/science.1199339},
  DOI = {10.1126/science.1199339},
  number = {6019},
  journal = {Science},
  publisher = {American Association for the Advancement of Science (AAAS)},
  author = {G\"{o}hler,  B. and Hamelbeck,  V. and Markus,  T. Z. and Kettner,  M. and Hanne,  G. F. and Vager,  Z. and Naaman,  R. and Zacharias,  H.},
  year = {2011},
  month = Feb,
  pages = {894–897}
}

@article{Mishra2013,
author = {Debabrata Mishra  and Tal Z. Markus  and Ron Naaman  and Matthias Kettner  and Benjamin Göhler  and Helmut Zacharias  and Noga Friedman  and Mordechai Sheves  and Claudio Fontanesi },
title = {Spin-dependent electron transmission through bacteriorhodopsin embedded in purple membrane},
journal = {Proceedings of the National Academy of Sciences},
volume = {110},
number = {37},
pages = {14872-14876},
year = {2013}
}

@article{Kettner2018,
author = {Kettner, Matthias and Maslyuk, Volodymyr V. and Nürenberg, Daniel and Seibel, Johannes and Gutierrez, Rafael and Cuniberti, Gianaurelio and Ernst, Karl-Heinz and Zacharias, Helmut},
title = {Chirality-Dependent Electron Spin Filtering by Molecular Monolayers of Helicenes},
journal = {The Journal of Physical Chemistry Letters},
volume = {9},
number = {8},
pages = {2025-2030},
year = {2018}
}

@article{gutierrez2023lanthanide,
  title={Lanthanide molecular nanomagnets as probabilistic bits},
  author={Guti{\'e}rrez-Finol, Gerliz M and Gim{\'e}nez-Santamarina, Silvia and Hu, Ziqi and Rosaleny, Lorena E and Cardona-Serra, Salvador and Gaita-Ari{\~n}o, Alejandro},
  journal={npj Computational Materials},
  volume={9},
  number={1},
  pages={196},
  year={2023},
  publisher={Nature Publishing Group}
}

@article{Fink1999,
  title = {Electrical conduction through DNA molecules},
  volume = {398},
  ISSN = {1476-4687},
  url = {http://dx.doi.org/10.1038/18855},
  DOI = {10.1038/18855},
  number = {6726},
  journal = {Nature},
  publisher = {Springer Science and Business Media LLC},
  author = {Fink,  Hans-Werner and Sch\"{o}nenberger,  Christian},
  year = {1999},
  month = apr,
  pages = {407–410}
}

@article{Nomura2019,
  title = {Phonon Magnetochiral Effect},
  author = {Nomura, T. and Zhang, X.-X. and Zherlitsyn, S. and Wosnitza, J. and Tokura, Y. and Nagaosa, N. and Seki, S.},
  journal = {Phys. Rev. Lett.},
  volume = {122},
  issue = {14},
  pages = {145901},
  numpages = {5},
  year = {2019},
  month = {Apr},
  publisher = {American Physical Society},
  doi = {10.1103/PhysRevLett.122.145901},
  url = {https://link.aps.org/doi/10.1103/PhysRevLett.122.145901}
}

@article{Shang2022,
author = {Shang, Zixuan and Liu, Tianhan and Yang, Qianqian and Cui, Shuainan and Xu, Kailin and Zhang, Yu and Deng, Jinxiang and Zhai, Tianrui and Wang, Xiaolei},
title = {Chiral-Molecule-Based Spintronic Devices},
journal = {Small},
volume = {18},
number = {32},
pages = {2203015},
doi = {https://doi.org/10.1002/smll.202203015},
url = {https://onlinelibrary.wiley.com/doi/abs/10.1002/smll.202203015},
eprint = {https://onlinelibrary.wiley.com/doi/pdf/10.1002/smll.202203015},
year = {2022}
}

@ARTICLE{Naaman1999,
  title     = "Asymmetric scattering of polarized electrons by organized
               organic films of chiral molecules",
  author    = "Ray, K and Ananthavel, S P and Waldeck, D H and Naaman, R",
  journal   = "Science",
  publisher = "American Association for the Advancement of Science (AAAS)",
  volume    =  283,
  number    =  5403,
  pages     = "814--816",
  month     =  feb,
  year      =  1999,
  language  = "en"
}

@article{Yan2024,
   author = "Yan, Binghai",
   title = "Structural Chirality and Electronic Chirality in Quantum Materials", 
   journal= "Annual Review of Materials Research",
   year = "2024",
   volume = "54",
   number = "Volume 54, 2024",
   pages = "97-115",
   doi = "https://doi.org/10.1146/annurev-matsci-080222-033548",
   url = "https://www.annualreviews.org/content/journals/10.1146/annurev-matsci-080222-033548",
   publisher = "Annual Reviews",
   issn = "1545-4118",
   type = "Journal Article",
  }

@article{GutierrezFinol2025DAISY,
      title={DAISY: simple simulation of spin dynamics in nanoscale magnets},
      DOI={10.26434/chemrxiv-2025-1118f}, journal={ChemRxiv},
      author={Gutiérrez-Finol, Gerliz Mercedes and Rosaleny, Lorena Estefanía and Camacho-Llovera, Amparo and Gaita-Ariño, Alejandro},
      year={2025}}

@ARTICLE{GutierrezFinol2026-CupFlow,
  title     = "A scalable kinetic Monte Carlo platform for charge transport
               dynamics in polymer-based memristive systems",
  author    = "Guti{\'e}rrez-Finol, Gerliz M and Zinovjev, Kirill and
               Gaita-Ari{\~n}o, Alejandro and Cardona-Serra, Salvador",
  journal   = "Mater. Chem. Front.",
  publisher = "Royal Society of Chemistry (RSC)",
  year      =  2026,
  copyright = "http://creativecommons.org/licenses/by-nc/3.0/",
  language  = "en"
}

@article{fransson2024current,
  title={Current induced spin-polarization in chiral molecules},
  author={Fransson, Jonas and Turin, L},
  journal={The Journal of Physical Chemistry Letters},
  volume={15},
  number={24},
  pages={6370--6374},
  year={2024},
  publisher={ACS Publications}
}

@article{rikken2022dielectric,
  title={Dielectric magnetochiral anisotropy},
  author={Rikken, Geert LJA and Avarvari, Narcis},
  journal={Nature Communications},
  volume={13},
  number={1},
  pages={3564},
  year={2022},
  publisher={Nature Publishing Group UK London}
}

@article{Meng2021,
  title = {Interface-driven electrical magnetochiral anisotropy in Pt/PtMnGa bilayers},
  volume = {118},
  ISSN = {1077-3118},
  url = {http://dx.doi.org/10.1063/5.0054662},
  DOI = {10.1063/5.0054662},
  number = {25},
  journal = {Applied Physics Letters},
  publisher = {AIP Publishing},
  author = {Meng,  K. K. and Chen,  J. K. and Miao,  J. and Xu,  X. G. and Jiang,  Y.},
  year = {2021},
  month = jun 
}
\vspace{3cm}

\end{document}